\documentclass[manuscript]{aastex}

\newcommand{\gray}{$\gamma$-ray\ } \newcommand{\grays}{$\gamma$-rays\ }
 
\newcommand{\etal}{et al.}

\begin{document}

\title{A Determination of the Intergalactic Redshift Dependent UV-Optical-NIR Photon Density Using
Deep Galaxy Survey Data and the Gamma-ray Opacity of the Universe}

\author{Floyd W.    Stecker}  
\affil{Astrophysics Science Division, NASA/Goddard  Space   Flight  Center}
\authoraddr{Greenbelt,     MD     20771;     Floyd.W.Stecker@nasa.gov}
\affil{Department of Physics  and Astronomy, University of California,
Los  Angeles}   
\authoraddr{Los  Angeles,  CA90095-1547}  
\author{Matthew A. Malkan}  
\affil{Department  of Physics  and  Astronomy, University  of
California,   Los  Angeles}  
\authoraddr{Los   Angeles,  CA90095-1547;
malkan@astro.ucla.edu}
\author{Sean T. Scully}  
\affil{Department  of Physics,  James  Madison
University} 
\authoraddr{Harrisonburg, VA 22807; scullyst@jmu.edu}

\begin{abstract}

We calculate the intensity and photon spectrum of the intergalactic background light (IBL) as
a function of redshift using an approach based on observational data obtained in many different
wavelength bands from local to deep galaxy surveys. This allows us
to obtain an empirical determination of the IBL and to quantify its 
observationally based uncertainties. Using our results on the IBL, we then place 68\% confidence
upper and lower limits on the opacity of the universe to $\gamma$-rays, free of
the theoretical assumptions that were needed for past calculations. We compare
our results with measurements of the extragalactic background light and upper limits obtained
from observations made by the {\it Fermi} ~Gamma-ray Space Telescope.

\end{abstract}

\subjectheadings{diffuse radiation -- galaxies:observations -- gamma rays:theory}

\pagebreak

\section{Introduction}

\subsection {Empirical Approach to Determining the Intergalactic Background Radiation}

The purpose of this paper is to present the results of a new, fully empirical approach to calculating the intergalactic background light (IBL) as well as the $\gamma$-ray opacity of the Universe. This methodology, hitherto unavailable, is now enabled by very recent data from deep galaxy surveys spanning the electromagnetic spectrum from millimeter to UV wavelengths and using galaxy luminosity functions for redshifts $ 0 \le z \le 8$ in the UV and for redshifts up to ~2
or ~3 in other wavelength ranges. We stress that this approach is both
capable of {\it delineating empirically based uncertainties} on the determination of the IBL, and the $\gamma$-ray opacity of the Universe.

In this paper (Paper I) we specifically
consider the frequency range from the far ultraviolet (FUV) to the near infrared $I$ band (NIR), as this range is of particular relevance to the \gray opacity studies in the $\sim$0.1-200 GeV energy range being made by the  {\it Fermi} \gray space telescope. A follow-up paper (Paper II) will address the frequency range from the NIR to the far-IR (FIR). That range has particular relevance for opacity studies by ground-based air \v{C}erenkov telescopes.

Previous calculations the IBL at different redshifts have been based on various theoretical models and assumptions. These include backward evolution models (Malkan \& Stecker 1998, 2001; Stecker, Malkan \& Scully 2006; Franceschini et al. 2008), 
semi-analytical forward evolution models ({\it e.g.}, Gilmore et al. 2009; Somerville et al. 2011) 
and other models based on the
evolution of galaxy parameters such as star formation rate and stellar population synthesis models
(Salamon \& Stecker 1998 (hereafter SS98); Kneiske et al. 2004). Kneiske \& Dole (2010) 
have recently used a forward evolution model to derive lower limits on the EBL. Finke, Razzaque \& Dermer (2010) employed a triple blackbody approximation to extimate the EBL. Dom\'{i}nguez et al. (2011) used an approach based on the redshift evolution of the $K$-band galaxy luminosity functions (LFs) derived by Cirasuolo et al. (2010), together with model templates based on {\it Spitzer}-based $0.2 \le z \le 1$ infrared galaxy SEDs and {\it AEGIS} data. To obtain $K$-band LFs for $1 <z <4$,  Cirasuolo et al. (2010) used 8 $\mu$m Spitzer/IRAC (Infrared Array Camera) channels combined with population synthesis models of Bruzual \& Charlot (2003), including a correction for dust obscuration. Most recently, a semi-analytic model of the EBL has been published by Gilmore et el. (2012). The earlier exploration of the EBL using direct measurements, galaxy counts, and indirect constraints was reviewed some time ago by Hauser \& Dwek (2001).

We note that previous studies had to adopt at least some assumptions about how galaxy LFs evolves with cosmic time, starting either at the present (well-measured epoch) and going back in time, or starting with the simulations of the galaxy formation epoch using semi-analytic models (see above) or modeled galaxy SEDs.  However, the latest observations have become sufficiently extensive and accurate to allow {\it direct integration} of observational data on galaxy LFs from the deep galaxy surveys at {\it many} wavelengths, where we can interpolate between observationally determined LFs at many wavelengths from the far UV to near infrared and the redshift range extending in the UV from $z = 0$ to $z \ge 8$.
Thus, the first goal of our paper is to determine the IBL based on empirical data
from deep survey galaxy observations. This 
avoids the complications entailed by theoretical calculations that have need of making
various assumptions for stellar  population  synthesis models, stellar initial mass
functions, unknown amounts of dust
extinction, and poorly known stellar metallicity-age  modeling for  different
evolving galaxy types ({\it e.g.}, Wilkins et al. 2012). This is because the observational data 
are the direct result of all of the physical processes involved in producing galactic emission.
Thus our treatment only involves uncertainties inherent in the analyses discussed in the
observational survey papers that we used.

\subsection{Gamma Ray Opacity and the IBL}

The second goal of our paper is to use our results on the IBL to determine the \gray opacity of the universe as a function of energy and redshift. 
It was first suggested by Stecker, De Jager \& Salamon (1992) 
that $\gamma$-ray observations from high redshift sources
such as blazars (and later $\gamma$-ray bursts) could be used to probe the IBL. Such  studies make  use of the  opacity caused by  the  annihilation  of   $\gamma$-rays  owing  to  
interactions  with  low energy  photons that produce $e^{+}e^{-}$ pairs. 
The {\it  Fermi Gamma-Ray Space Telescope (Fermi)} is now being used to probe the high
redshift IBL at optical and UV  wavelengths by constraining the opacity
of the universe to multi-GeV $\gamma$-rays (Abdo et al. 2010).
This is accomplished by measuring the energy of the highest energy
photons observed by {\it Fermi} that have been emitted by GRBs and 
blazars at known redshifts.

Observations of TeV  $\gamma$-ray emitting  blazars utilizing
modern air \v{C}erenkov  telescope arrays  also  probe, or at least constrain, the
nearby  (redshift  $z \sim  0-0.5$) intergalactic  infrared  background
radiation. Attempts to constrain the IBL have been made by various authors 
(Stecker \& de Jager 1993; Aharonian et al. 2006 (but see Stecker, Baring \& Summerlin 2009); Mazin \& Raue 2007; Georganopoulos, Fincke \& Reyes 2010; Abdo et al. 2010; Orr, Krennrich \& Dwek 2011, but see Stecker, Baring \& Summerlin 2009).

Our methodology will also be used to define secure upper and lower limits on the opacity of the universe to high energy $\gamma$-rays based on the observational uncertainties in the deep survey data. We then compare the opacity range defined by these limits with the upper limits derived using the {\it Fermi} observations of multi-GeV $\gamma$-rays from high redshift sources Abdo, et al (2010).

\section{Intergalactic Photon Energy Densities and Emissivities from Galaxies }

The co-moving radiation energy density $u_{\nu}(z)$ is
derived from the co-moving specific emissivity ${\cal E}_{\nu}(z)$,
which, in turn is derived from the galaxy luminosity function (LF). The galaxy luminosity 
function,~$\Phi_{\nu}(L)$, is defined
as the distribution function of galaxy luminosities at a specific frequency
or wavelength. The specific emissivity at frequency $\nu$ and redshift $z$ 
(also referred to in the literature as the
luminosity density, $\rho_{L_{\nu}}$) , is the integral over the luminosity function

\begin{equation}
 {\cal E}_{\nu}(z) = \int_{L_{min}}^{L_{max}} dL_{\nu} \, L_{\nu} \Phi(L_{\nu};z)
\label{phi}
\end{equation}

There are many references in the literature where the LF is given and fit to Schechter
parameters, but where $\rho_{L_{\nu}}$ is not given. In those cases, we could not determine
the covariance of the errors in the Schechter parameters used to determine the dominant statistical errors in their analyses. Thus, we could not ourselves accurately determine
the error on the emissivity from equation (\ref{phi}). We therefore chose to use only the
papers that gave values for $\rho_{L_{\nu}}(z) = {\cal E}_{\nu}(z)$ with errors. We did not consider cosmic variance, but this uncertainly should be minimized since we used data from many surveys.

In compiling the observational data on ${\cal E}_{\nu}(z)$, we scaled all of the results to 
a value of h = 0.7. Thus results using h = 0.5 were scaled by a factor of (7/5)\footnote{Using
the most recent and accurate value of 0.74 (Riess et al. 2011) would increase all of our results by $\sim 6\%$}. 

The co-moving radiation energy density $u_{\nu}(z)$ 
is the
time integral of the co-moving specific emissivity ${\cal E}_{\nu}(z)$,
\begin{equation} 
\label{u1}
u_{\nu}(z)=
\int_{z}^{z_{\rm max}}dz^{\prime}\,{\cal E}_{\nu^{\prime}}(z^{\prime})
\frac{dt}{dz}(z^{\prime})e^{-\tau_{\rm eff}(\nu,z,z^{\prime})},
\end{equation}

\noindent where $\nu^{\prime}=\nu(1+z^{\prime})/(1+z)$ and $z_{\rm max}$ is the
redshift corresponding to initial galaxy formation 
(Salamon \& Stecker 1998, hereafter SS98), and
\begin{equation}
\frac{dt}{dz}{(z)} = {[H_{0}(1+z)\sqrt{\Omega_{\Lambda} + \Omega_{m}(1+z)^3}}]^{-1},
\label{cosmology}
\end{equation}

\noindent with $\Omega_{\Lambda} = 0.72$ and $\Omega_{m} = 0.28$.

The opacity factor for frequencies below the Lyman limit is dominated by dust extinction.
In the model of SS98, which relied on the population synthesis studies
of Bruzual \& Charlot (1993), dust absorption was not included. Our earlier paper
(Stecker, Malkan \& Scully 2006) used a rough approximation of the results obtained by
Salamon \& Stecker (1998) (SS98) and therefore,
also did not take dust absorption into account. However, since we are here using actual
observations of galaxies rather than models, dust absorption is implicitly included.
The remaining opacity $\tau_{\nu}$ refers to the extinction of ionizing photons
with frequencies above the rest frame Lyman limit of $\nu_{LyL} \equiv 3.29 \times 10^{15}$ Hz by interstellar and intergalactic hydrogen and helium. It has been shown that this opacity is 
very high, corresponding to the expectation of very small fraction of ionizing radiation in
intergalactic space compared with radiation below the Lyman limit (Lytherer \etal
1995; SS98). In fact, the Lyman limit cutoff is used as a tool; when galaxies disappear when
using a filter at a given waveband ({\it e.g.}, "$U$-dropouts", "$V$-dropouts") it is an indication of the redshift of the Lyman limit. We thus replace equation (\ref{u1}) with the following expression

\begin{equation} 
\label{u2}
u_{\nu}(z)=
\int_{z}^{z_{\rm max}}dz^{\prime}\,{\cal E}_{\nu^{\prime}}(z^{\prime})
\frac{dt}{dz}(z^{\prime}){{\cal H}(\nu(z') - \nu'_{LyL})},
\end{equation}

\noindent where ${\cal H}(x) $ is the Heavyside step function.


\subsection{Empirical Specific Emissivities}

\subsubsection{Luminosity Densities}

We have used the results of many galaxy surveys to compile a set of luminosity
densities, $\rho_{L_{\nu}}(z) = {\cal E}_{\nu}(z)$ (LDs), at all observed redshifts, and at rest-frame wavelengths from the far-ultraviolet, FUV = 150 nm  to the $I$ band, $I$ = 800 nm.  
Figure 1 shows the redshift evolution of the luminosity ${\cal E}_{\nu}(z)$ for the various wavebands based on those published in the literature.\footnote{Table 1 references used to construct Figure 1 are as follows: Bouwens \etal ~(2007)(BO07), Bouwens \etal ~(2010)(BO10), Budav\'{a}ri \etal (2005)(BU05), Burgarella \etal ~(2007)(BU07), Chen \etal (2003)~(CH03), Cucciati \etal ~(2012)(CU12), Dahlen \etal ~(2007)(DA07), Faber \etal ~(2007)(FA07) and references therein, Iwata \etal ~(2007)(IW07), Ly \etal ~(2009)(LY09), Reddy \& Steidel (2009)(RE09),   Marchesini \etal ~(2007)(MA07), Marchesini \& Van~ Dokkum 2007 (MAV07), Marchesini \etal ~(2012)(MA12), Oesch \etal ~(2010)(OE10), Paltani \etal~ (2007)(PA07), Reddy ~\etal ~(2008)(RE08), Sawicki \& Thompson (2006)(SA06), Schiminovich \etal ~(2005)(SC05), Steidel \etal ~(1999)(ST99), Tresse \etal ~(2007) (TR07), Wolf \etal (2003) (WO03), Wyder \etal ~(2005)(WY05), Yoshida \etal ~(2006)(YO06).}
The lower right panel shows all of the observational determinations of galaxy LDs from the references in footnote 2. 
The specific waveband and mean redshift identifications for these data are listed in Table 1 using the key abbreviations indicated in footnote 2. 
This table reflects the fact that direct determinations of galaxy LDs are only available out to 
an {\it observed wavelength} of about 2.2 $\mu$m (rest wavelength $2.2/(1+z) ~\mu$m). This is because any attempt to survey large areas of the sky with ground-based telescopes in wavebands longer than 2$\mu$m is prevented by the sudden increases in background noise.\footnote{This 2$\mu$m barrier is only circumvented by using space-based mid-infrared (3 to 8$ \mu$m) telescopes such as
{\it AKARI} (with its Infrared Camera, IRC), and {\it Spitzer} (with its Infrared Array Camera, IRAC). These telescopes have only conducted multi-band imaging and redshift surveys with the necessary sensitivity to measure the high-redshift ($z \ge 2$) galaxy population in a few, relatively small deep fields.}

Thus, at redshifts above 1.6, the longest rest-wavelengths under consideration no longer have well measured LDs. 
At these longer wavelengths, we are obliged to fall back on a secondary method for estimating galaxy luminosities: we use the closest available LDs,
and extrapolate them using the average observed {\it color} of galaxies from measurements at that redshift.  This 'minimal extrapolation'
should be reliable because the average galaxy colors, especially at long wavelengths, change only gradually with redshift.  For example, the galaxies that are included in the rest-frame $R$ band LD at $z = 2.2$ by Marchesini et al. are very similar to those of the galaxies that would have been included in  an $I$-band LD at that redshift.  Since we are only extrapolating by a small step in wavelength ($\Delta \lambda/\lambda \sim 0.15$), it is quite reasonable to shift the $R$-band LD using the average $R - I$ colors observed at that redshift.  The incremental color shifts we apply become large only at $z \ge 4$, where, as we show in Section 4, the overall contributions to the IBL $\gamma$-ray opacity are not very substantial. 
Our color relations, which are also used to interpolate between the closely spaced wavebands, are given the next subsection. They are given as a function of redshift, $z$, since galaxies tend to be bluer on average at higher redshifts.  

\subsubsection{Average Colors}

It is hardly surprising that there are often large apparent jumps, or changes, in the shape and the normalization of the LDs going from one waveband to an immediately adjacent one. We therefore applied an independent test of the consistency of these LDs, by comparing the integrated {\it ratios} of LDs at adjacent wavebands to the published average colors measured by observers. This test has the great advantage of not requiring accurate estimates of volume incompleteness or even very accurate redshifts.  Broadband colors ({\it i.e.,} local continuum slopes) are easier to measure than LDs.  The main problem is that all galaxy samples at all redshifts show a wide observed range of broadband colors. The typical $1 \sigma$ scatter we found in published color distributions was $\pm~ 0.5$ mag. A few rest-frame colors that are very sensitive to stellar population, such as $U-B$, often show even larger variation. 

In order to determine the redshift evolution of the LD in each of the bands out to a redshift of $\sim$ 8, we utilized color relations to transform data from other bands. We have chosen to include all data possible in excess of $z = 1.5$ to fill in the gaps for various wavebands mostly at higher redshifts.\footnote{The most comprehensive observations of galaxies in the best observed Deep Fields include extremely sensitive Spitzer/IRAC photometry. The IRAC data are most complete
in its Band 1 (3.6 $\mu$m observed) wavelength, and gradually become less sensitive out
to the reddest IRAC band at 8 $\mu$m observed wavelength corresponding to a rest wavelength of
$8/(1+z) ~\mu$m.}. This also provides both an overlap to existing data and multiple sources of data as a check for consistency of our color relations. 

Published estimates of {\it average colors} from galaxy surveys at various wavebands
and redshifts tend to be bluer at shorter wavelengths, and redder at longer wavelengths.  This is due to the composite nature of stellar populations in galaxies, with hot young stars making a stronger contribution in the UV portion of the spectrum while red giants dominate the long wavelengths. Thus, the galaxies that are included in a UV LF and not all the same galaxies as those included in an LF in the $R$ band. 

There is a clear trend with redshift over all wavelengths, which is well known. Redder galaxies ({\it e.g.}, local E and S0 galaxies) are more and more outnumbered by blue, actively star-forming galaxies, at higher redshifts. The average characteristic age of stellar populations decreases with redshift. Our color relations agree with this trend. At the highest redshifts most known galaxies are dominated by young starburst populations of O and B stars. This tends to produce very blue overall spectral energy distribution without very much sensitivity to the exact details of the star formation. These factors are automatically taken into account when one
uses the actual observational data on the LDs at various wavelengths and redshifts.

Defining the average wavelengths of the various bands in $nm$ as follows:

\noindent FUV = 150, NUV = 280, $U$ = 365, $B$ = 445, $V$ = 551, $R$ = 658, $I$ = 806 nm

We then use the commonly measured astronomical 
parameter $\beta$, which is defined by the relation between the differential flux and wavelength
of a galaxy, $f_{\lambda} \propto \lambda^{\beta}$.  We have adopted the following relations (colors) for $\beta_{\Delta\lambda}(z)$:

\vspace{5pt}
$\beta (FUV - NUV) = -1.0 - 1.25 log (1 + z), \ log (1 + z) \le 0.8 $

\noindent derived from Bouwens, et al. (2009); Budav{\'a}ri et~al.(2005); Castellano et al. (2012); Cucciati, et al. (2012); Dunlop et al. (2012); Willott, et al. (2012); Wyder et~al.(2005),

\vspace{5pt}
$\beta (B - V) = +0.3 - 1.6 log (1 + z), \ log (1 + z) \le 0.6 $

\noindent derived from Arnouts et~al.(2007); Brammer (2011),

\vspace{5pt}
$\beta(NUV - U) = +0.5 - 1.2 log (1 + z), \ log (1 + z) \le 0.6 $

\noindent derived from Tresse et al. (2007),

\vspace{5pt}$
\beta (NUV - R) = +2.5 - 6.0 log (1 + z), \ log (1 + z) \le 0.6$

\vspace{5pt}
$\beta (U - V) = +1.3 - 3.0 log (1 + z), \ log (1 + z) \le 0.6 $

\noindent derived from Arnouts, et al. (2007); Brammer (2011): Ly et al. (2009),

\vspace{5pt}
$\beta (U - B) = +3.0 - 5.0 log (1 + z), \ log (1 + z) \le 0.6 $

\noindent derived from Marchesini et al. (2007); Gonz\'{a}lez et al. (2011),
\vspace{5pt}

\noindent For the FUV-NUV relation we set $\beta [log (1+z) > 0.8] = \beta [log (0.8)].$ For all of the other relations we set $\beta [log (1+z) > 0.6] = \beta [log (0.6)]. $

We used the above redshift-dependent relations where appropriate in our analysis. We stress that in the redshift ranges
where they overlap, {\it the colored (observational) data points shown for the various wavelength bands in Figure 1 agree quite well, within the uncertainties, with the black data points that were extrapolated from the shorter wavelength bands using our color relations.} Also, where there is
no overlap at the higher redshifts, the uncertainty bands in photon density (see next section) show no discontinuities.

\subsection{Photon Density Calculations}

The observationally determined LDs, combined with the color relations, extend our coverage of galaxy photon production from the FUV to the NIR in the galaxy rest frame.  We have at least one or two determinations at each wavelength across the most crucial redshift range $0 \le z \le 2.5$.
However, to calculate the opacity for photons at energies higher than $\sim 250/(1+z)$ GeV
(see next section), requires the determination of galaxy LDs
at longer rest wavelengths and higher redshifts.  These regimes are less
well constrained by observations, since they require measurement of
very faint galaxies at long wavelengths (mid-IR observed frame.) We will address this topic further in Paper II. We have assumed a constant color at high redshift at the longer wavelengths as stated above. However, we stress that our final results are not very sensitive to errors in our average color relations because the interpolations that we make cover very small fractional wavelength intervals, 
$\Delta\lambda(z)$. We have directly tested this by numerical trial.

The second goal of our paper is to place upper and lower limits (within a 68\% confidence band)
on the opacity of the universe to \grays . These limits are a direct result of the 68\% confidence band upper and lower limits of the IBL determined from the observational data on $\rho_{L_{\nu}}$ . In order to determine these limits, we make no assumptions about the luminosity density evolution.  We derive a luminosity confidence band in each waveband by using a robust rational fitting function characterized by
\begin{equation}
\rho_{L_{\nu}} = {\cal E}_{\nu}(z) = {{ax+b}\over{cx^{2}+dx+e}}
\label{rational}
\end{equation}
where $x = \log(1+z)$ and $a$,$b$,$c$,$d$,and $e$ are free parameters.

The 68\% confidence band is then computed
from Monte Carlo simulation by finding 100,000 realizations of the data and then fitting the rational function. In order to best represent the tolerated confidence band, particularly at the highest redshifts, we have chosen to equally weight all FUV points in excess of a redshift of 2.  Our goal is not to find the best fit to the data but rather the limits tolerated by the current observational data.  In order to perform the Monte Carlo of the fitting function, a likelihood is determined at each redshift containing data.  The shape of the function is taken to be Gaussian (or the sum of Gaussians where multiple points exist) for symmetric errors quoted in the literature.  Where symmetric errors are not quoted it is impossible to know what the actual shape of the likelihood functions is.  We have chosen to utilize a skew normal distribution to model asymmetric errors.   This assumption has very little impact on the determination of the confidence bands. The resulting bands are shown along with the luminosity density data in Figure 1.

With the confidence bands established, we take the upper and lower limits of the bands to be our high and low IBL respectively.  We then interpolate each of these cases separately  between the various wavebands to find the upper and lower limit rest frame luminosity densities.  The calculation is extended to the Lyman limit using the slope derived from our color relationship between the near and far UV bands.

The specific emissivity is then the derived high and low IBL luminosity densities ${\cal E}_{\nu}(z) = \rho_{L_{\nu}}(z)$. The co-moving radiation energy density is determined from equation \ref{u2}.  Figure 2 shows the resulting photon density determined by dividing the energy density by the energy in each frequency for high and low IBL.  This result is used as input for the determination of the optical depth of the universe to \grays . 

The photon densities
\begin{equation}
\epsilon n(\epsilon,z) = u(\epsilon,z)/\epsilon \ \ ,
\label{gammadens}
\end{equation}
with $\epsilon = h\nu$, as
calculated using equation (\ref{u1}), are shown in Figure 2.

\section{Comparison of z = 0 IBL with Data and Constraints}

As a byproduct of our determination of the IBL as a function of redshift using LDs from galaxy surveys, we have also determined the local ($z = 0$) IBL, also known as the extragalactic background light (EBL). Determining the EBL directly has been the object of intense observational effort,
although the various estimates and limits in the published literature are far from consistent with each other. Nonetheless, since these
observations provide a potential consistency check on our calculations, we consider them here.   

Using equation (\ref{u1}), together with our empirically based determinations given the confidence band derived for our specific emissivities, ${\cal E}_{\nu}(z)$, we have evaluated the EBL within the 68\% confidence band upper and lower limits within the wavelength range of our calculations. This band is indicated by the gray zone in Figure 3. We also show recent measurements using the Hubble Wide-field Planetary Camera 2 (Bernstein 2007), the dark field from Pioneer 10/11 (Matsuoka et al. 2011) and the preliminary analysis of Mattila et al. (2011) using differential measurements using the ESO VLT (very large telescope array).  Figure 3 also shows the various lower limits
from galaxy counts obtained by Gardner et al. (2000) from the ST Imaging Spectrograph data,
by Madau \& Pozzetti (2000) using Hubble Deep Field South data, and by Xu et al. (2005) from GALEX (Galaxy Evolution Explorer) data, all indicated by upward-pointing arrows. 


\section{The Optical Depth from Interactions with Intergalactic Low Energy Photons}

The cross section for photon-photon scattering to electron-positron pairs can be calculated using quantum electrodynamics (Breit \& Wheeler 1934).  
The threshold for this interaction is determined from the frame invariance of the square of the four-momentum vector that reduces to the square of the threshold energy, $s$, required to produce twice the electron rest mass in the c.m.s.:  
\begin{equation}
s = 2\epsilon E_{\gamma} (1-\cos\theta) = 4m_{e}^2
\label{s}
\end{equation}

This invariance is known to hold to within one part in $10^{15}$ (Stecker \& Glashow 2001;
Jacobson, Liberati, Mattingly \& Stecker 2004).

With the co-moving energy density $u_{\nu}(z)$ evaluated, the optical
depth for \grays owing to electron-positron pair production 
interactions with photons of the stellar radiation
background can be determined from the expression (Stecker, De Jager, \& Salamon ~1992)

\begin{equation} 
\label{tau}
\tau(E_{0},z_{e})=c\int_{0}^{z_{e}}dz\,\frac{dt}{dz}\int_{0}^{2}
dx\,\frac{x}{2}\int_{0}^{\infty}d\nu\,(1+z)^{3}\left[\frac{u_{\nu}(z)}
{h\nu}\right]\sigma_{\gamma\gamma}[s=2E_{0}h\nu x(1+z)],
\label{tau}
\end{equation}

~
In equations (\ref{s}) and  
(\ref{tau}), $E_{0}$ is the observed \gray energy at redshift zero, 
$\nu$ is the frequency at
redshift $z$,
$z_{e}$ is the redshift of
the \gray source at emission, $x=(1-\cos\theta)$, \\
$\theta$ being the angle between
the \gray and the soft background photon, $h$ is Planck's constant, and
the pair production cross section $\sigma_{\gamma\gamma}$ is zero for
center-of-mass energy $\sqrt{s} < 2m_{e}c^{2}$, $m_{e}$ being the electron
mass.  Above this threshold, the pair production cross section is given by

\begin{equation} 
\label{sigma}
\sigma_{\gamma\gamma}(s)=\frac{3}{16}\sigma_{\rm T}(1-\beta^{2})
\left[ 2\beta(\beta^{2}-2)+(3-\beta^{4})\ln\left(\frac{1+\beta}{1-\beta}
\right)\right],
\end{equation} 

\noindent where $\sigma_T$ is the Thompson scattering cross section and $\beta=(1-4m_{e}^{2}c^{4}/s)^{1/2}$  (Jauch \& Rohrlich 1955).

It follows from equation (\ref{s}) that the pair-production cross section energy has a threshold at $\lambda = 4.75 \ \mu {\rm m} \cdot E_{\gamma}({\rm TeV})$. 
\noindent Since the maximum $\lambda$ that we consider
here is in the rest frame I band at 800 nm at redshift $z$, and we observe $E_{\gamma}$ at redshift 0, so that its energy at
interaction in the rest frame is $(1+z)E_{\gamma}$, we then get a
conservative upper limit on $E_{\gamma}$ of $\sim 200(1+z)^{-1}$ GeV as the maximum \gray energy
affected by the photon range considered here. Allowing for a small error, our opacities
are good to $\sim 250(1+z)^{-1}$ GeV.
The 68\% opacity ranges for $z = 0.1,0.5, 1, 3 ~$and $5$, calculated using equation (\ref{tau})
are plotted in Figure 4. 

The widths of the grey uncertainty ranges in the LDs shown in Figure 1 increase towards higher redshifts, especially at the longest rest wavelengths.  This reflects
the decreasing amount of long-wavelength data and the corresponding increase in
uncertainties about the galaxies in those regimes. However,
these uncertainties do not greatly influence
the opacity calculations. Because of the short time interval of the emission from
galaxies at high redshifts their photons do not contribute greatly to the opacity at lower redshifts. Indeed, Figure 4 shows that the opacities determined for redshifts of 3 and 5 overlap
within the uncertainties.
\section{Results and Implications }

We have determined the IBL using local and deep galaxy survey data, together with observationally produced uncertainties, for wavelengths from 150 nm to 800 nm and redshifts out to $z > 5$. We have presented our results in terms of  68\% confidence band upper and lower limits. 
As expected, our $z = 0$ (EBL) 68\% lower limits are higher than those obtained by galaxy counts alone, since the EBL from galaxies is not completely resolved.
Our results are also above the theoretical lower limits given recently by
Kneiske and Dole (2010). In Figure 3, we compare our $z = 0$ result with both published and preliminary measurements and limits.

Figure 5 shows our 68\% confidence band for $\tau = 1$ on an energy-redshift plot
compared with the {\it Fermi} data on the highest energy photons from extragalactic
sources at various redshifts as given by Abdo et al. (2010). It can be seen that none of the
photons from these sources would be expected to be significantly annihilated by pair
production interactions with the IBL. This point is brought out further in Figure 6.
This figure compares the 68\% confidence band of our opacity results with the 95\% confidence
upper limits on the opacity derived for specific blazars by Abdo et al. (2010).

For purposes of discussion, we mention some points of comparison with previous work.
Our EBL results for $z = 0$, while lower than the fast evolution model of our previous work, are generally higher than those modeled more recently. As an example, at a wavelength of 200 nm in the FUV range our uncertainty range is a 
factor of 1.8 - 4.2 higher than the recent fiducial semi-analytic model of Gilmore et al. (2012) and similarly higher than the previous model result of Dominguez et al. (2011).
Our opacity results at $z \simeq 1$ are comparable to, or lower than, the models of Kneiske et al. (2004). They are also consistent with the results of the non-metallicity corrected model of SS98. However, they are higher than the models of Franceschini et al. (2008),
Gilmore et al. (2009), and Finke et al. (2010), as indicated by comparing Figure 3 of Abdo et al.
(2010) with our Figure 5. We stress that these comparisons are for illustrative purposes only.
Because our new methodology is based on the direct use of luminosity densities derived directly from
observations, we take the position that they stand by themselves and should be compared primarily
with the observational data as shown in our Figures 3, 5 and 6. In that regard, we find full
consistency within our observationally determined uncertainties.\footnote{While we were preparing our revised manuscript for publication a similar empirically based
calculation by Helgason \& Kashlinsky (2012) appeared on the arXiv. These authors calculated the 
EBL and \gray opacity based on galaxy luminosity functions compiled by Helgason, Ricotti \&
Kashlinsky (2012) extrapolated to $z \ge 2$ using an exponential cutoff in $z$. Their opacity results are generally consistent with the results presented here.}

Our result bears on questions regarding the
possible modification of the pair-production opacity effect on the \gray flux from distant
extragalactic sources, either by line-of-sight photon-axion oscillations during propagation 
({\it e.g.}, De Angelis et al. 2009) or by the addition of a component of secondary \grays from
interactions of blazar-produced cosmic-rays with photons along the line-of-sight to the
blazar ({\it e.g.},~Essey et al. 2010; Essey \& Kusenko 2012). Future theoretical studies and future \gray observations of extragalactic sources with {\it Fermi} and the {\it \v{C}erenkov Telescope Array}, which will be sensitive to extragalactic sources at energies above 10 GeV (Gernot 2011), should help to clarify these important aspects of high energy astrophysics.

\section{Our Results Online}

Our results in numerical form are available at the following link:

\noindent {\bf http://csma31.csm.jmu.edu/physics/scully/opacities.html}

\section*{Acknowledgments}

We would like to thank Luis Reyes and Anita Reimer for supplying us with the Fermi results shown in Figure 5. We thank Richard Henry for a helpful discussion of the UV background data. We also thank Tonia Venters for helpful discussions. This research was partially supported by a NASA Astrophysics Theory Grant and a NASA Fermi Guest Investigator Grant.



\section*{\small Table 1. Identification of References for Fig. 1 Data by Waveband and Redshift}
\tiny
 \begin{flushleft}   
  \begin{tabular}{@{\hspace{-1.0cm}}cccccccc}
    \hline
    z & FUV & NUV & U & B & V & R & I \\ 
     \hline
	.05 & SC05, WY05 & WY05 & ¥ & ¥ & ¥ & ¥ & ¥ \\
	.1 & BU05,CU12 & BU05,CU12 & ¥ & ¥ & ¥ & ¥ & ¥ \\
    .15 & TR07 & TR07 & TR07 & TR07 & TR07 & TR07 & TR07 \\ 
	.20 & BU05 & BU05 & ¥ & ¥ & & ¥ & ¥ \\
	.25 & WO03 & WO03 & ¥ & ¥ & & WO03 & ¥ \\  
    .3 & SC05,CU12,SC05,TR07 & TR07,CU12 & TR07,DA05 & TR07,DA05,FA07 & TR07 & TR07 & TR07 \\
	.35 &  & DA07, WO03 & ¥ & WO03 & ¥ & WO03 & ¥ \\
	.45 &  & WO03 & DA05 & DA05, WO03 & ¥ & WO03 & ¥ \\ 
    .5 & SC05, CU12, TR07 & TR07 & TR07 & TR07, FA07 & TR07 & TR07 & TR07 \\ 
    .55 & ¥ & DA07, WO03 & ¥ & WO03 &  & WO03 & ¥ \\
    .6 & ¥ &   & DA05 & DA05 &   & CH03 & ¥ \\
    .65 &  & WO03 &   & WO03 & MA12 & WO03 & ¥ \\ 
    .7 &   & TR07,CU12 & TR07 & TR07, FA07 & TR07 & TR07 & TR07 \\
    .75 &   & WO03 & ¥ & WO03 & ¥ & WO03 & ¥ \\
    .85 &   & WO03 & ¥ & WO03 & ¥ & WO03 & ¥ \\ 
    .9 & TR07,CU12 & TR07,CU12 & TR07, DA05 & TR07, DA05, FA07 & TR07 & TR07 & TR07 \\
    .95 &   & WO03 & DA05 & WO03, DA05 & MA12 & WO03, DA05 & ¥ \\ 
    1.0 & SC05 & WO03 & ¥ & WO03 & ¥ & WO03 & ¥ \\
    1.1 & CU12, TR07, DA07, BU07 & DA07,TR07,CU12, WO03 & TR07 & TR07, FA07, WO03 & TR07 & TR07, WO03 & TR07 \\
    1.2 &   &  & DA05 & DA05 &  & CH03, DA05 & \\
	1.3 & CU12, TR07 & TR07 & TR07 & TR07 & TR07 & TR07 & TR07 \\
	1.4 & CU12 & CU12 &  &  &  &   &  \\ 
    1.5 &  &  & DA05 & DA05 &   & DA05  &   \\
    1.6 & TR07 & TR07 & TR07 & TR07 & TR07 & TR07 & TR07 \\ 
	1.7 &  &  & DA05 & DA05 &   & DA05  &   \\ 
    1.8 & DA07 & DA07 & ¥ & ¥ & MA12 & ¥ & ¥ \\
    1.9 &  &  & DA05 & DA05 &   & DA05  &   \\
    2.0 & SC05 & ¥ & ¥ & ¥ & ¥ & ¥ & ¥ \\
    2.1 & CU12 & CU12 & ¥ & ¥ & ¥ & ¥ & ¥ \\
	2.2 & RE08, SA06 &  & ¥ & MA07 & MA07 & MA07 & ¥ \\ 
	2.3 & LY09 &  & ¥ &  &  &  & ¥ \\ 
    2.4 &  & ¥ & ¥ & ¥ & MA12 & ¥ & ¥ \\
    2.9 & SC05 & ¥ & ¥ & ¥ &  & ¥ & ¥ \\ 
    3.0 & CU12 & CU12 & ¥ & MA07 & MA07, MA12 & ¥ & ¥ \\ 
    3.5 & PA07 & ¥ & ¥ & ¥ & ¥ & ¥ & ¥ \\
    3.8 & BO07  & ¥ & ¥ & ¥ & MA12 & ¥ & ¥ \\ 
    4.0 & YO06,CU12 &   & ¥ & ¥ & ¥ & ¥ & ¥ \\
    4.1 & SA06  &   & ¥ & ¥ & ¥ & ¥ & ¥ \\
	4.8 & IW07 & ¥ & ¥ & ¥ & ¥ & ¥ & ¥ \\ 
    5.0 & BO07 & ¥ & ¥ & ¥ & ¥ & ¥ & ¥ \\
	5.9 & BO07 & ¥ & ¥ & ¥ & ¥ & ¥ & ¥ \\ 
    6.8 & BO11 & ¥ & ¥ & ¥ & ¥ & ¥ & ¥ \\
    7.0 & OE10 & ¥ & ¥ & ¥ & ¥ & ¥ & ¥ \\
	8.2 & BO10 & ¥ & ¥ & ¥ & ¥ & ¥ & ¥ \\	 
     \hline
  \end{tabular}
  \label{tab:label}
  \end{flushleft}

\newpage

\normalsize

\centerline{Figure Captions}

\noindent Figure 1: The observed specific emissivities in our fiducial wavebands.  The lower right panel shows all of the observational data from the references in footnote 1. In the other panels, non-band data have been shifted using the color relations
given in the text in order to fully determine the specific emissivities in each waveband. The symbol designations are FUV: black filled circles, NUV: magenta open circles, $U$: green filled squares, $B$: blue open squares, $V$: brown filled triangles, $R$: orange open triangles, 
$I$: yellow open diamonds.  Grey shading: 68\% confidence bands (see text).

\vspace{10pt}

\noindent Figure 2: The photon densities $\epsilon n(\epsilon)$ shown as a continuous function of photon energy and redshift for both the high (upper panel) and low (lower panel) IBL.

\vspace{10pt}

\noindent Figure 3: Our empirically-based determination of the EBL together with lower limits and data as described in the text. The legend is as follows: Madau \& Pozzetti(2000):Black Cicles,
Xu et al.(2005):Crosses, Gardner et al.(2000):Open Squares, Matsuoka et al.(2011):Open Circles,
Mattilla et al.(2011)(preliminary):Black Squares, Bernstein(2007):Black Diamonds. The upper limit
from Mattilla et al.(2011) is thickened for clarity.

\vspace{10pt}

\noindent Figure 4: Our empirically determined opacities for redshifts of 0.1, 0.5, 1, 3, 5.
The dashed lines are for $\tau = 1$ and $\tau = 3$.

\vspace{10pt}

\noindent Figure 5: A $\tau = 1$ energy-redshift plot (Fazio \& Stecker 1970) showing our uncertainty band results compared with the {\it Fermi} plot of their highest energy photons from FSRQs (red), BL Lacs (black) and and GRBs (blue) {\it vs.} redshift (from Abdo et al. 2010).

\vspace{10pt}

\noindent Figure 6: Our opacity results for the redshifts of the blazars compared with 95\% confidence opacity upper limits (red arrows) and 99\% confidence limits (blue arrows) as given by the {\it Fermi} analysis of Abdo, et al. (2010).

\clearpage


\begin{figure}
\begin{center}
\includegraphics[height=7in]{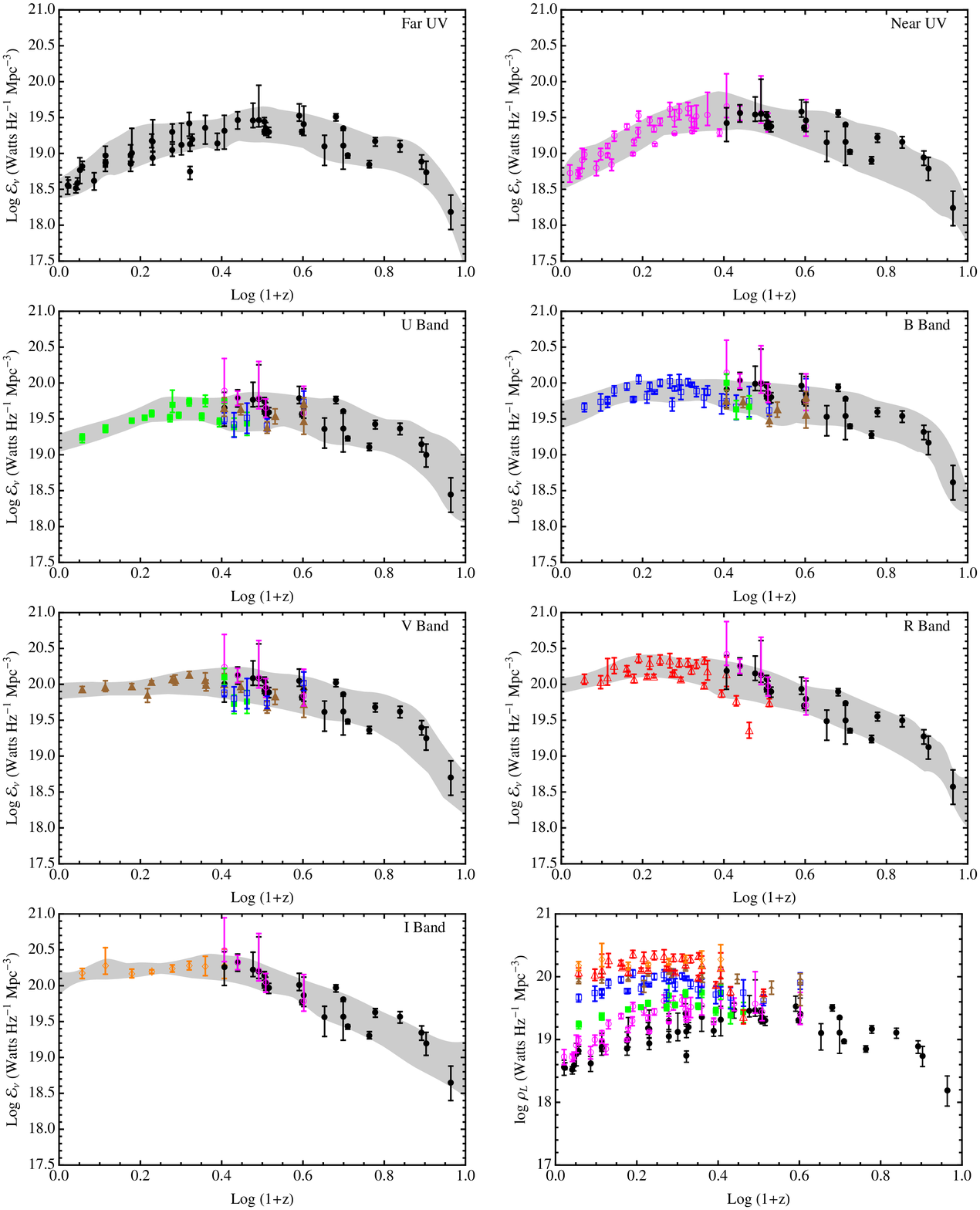}
Figure 1.
\label{confband}
\end{center}
\end{figure}

\clearpage

\begin{figure}
\begin{center}
\includegraphics[height=7in]{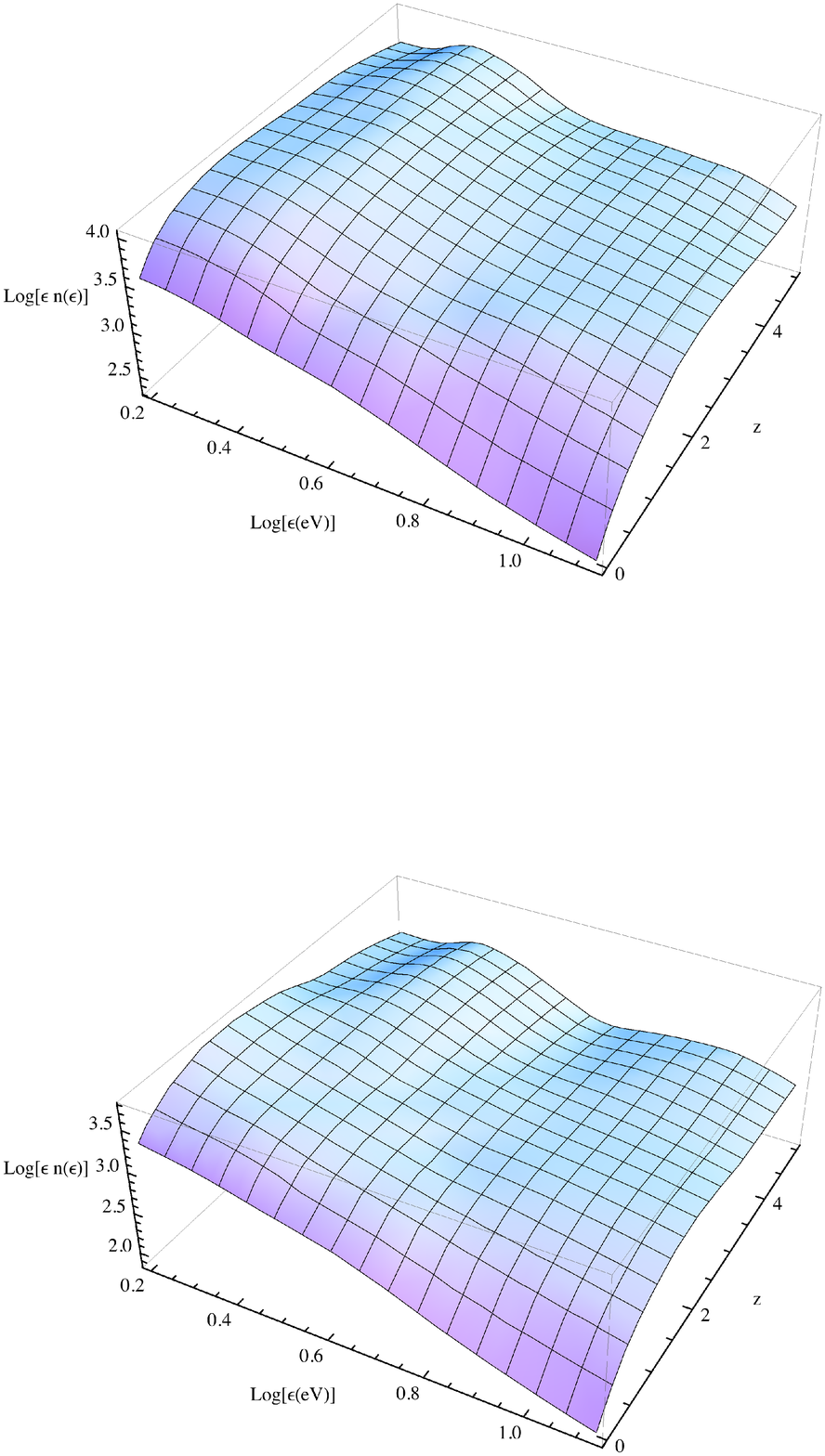}
Figure 2.
\label{dens}
\end{center}
\end{figure}

\clearpage

\begin{figure}
\begin{center}
\includegraphics[width=7in]{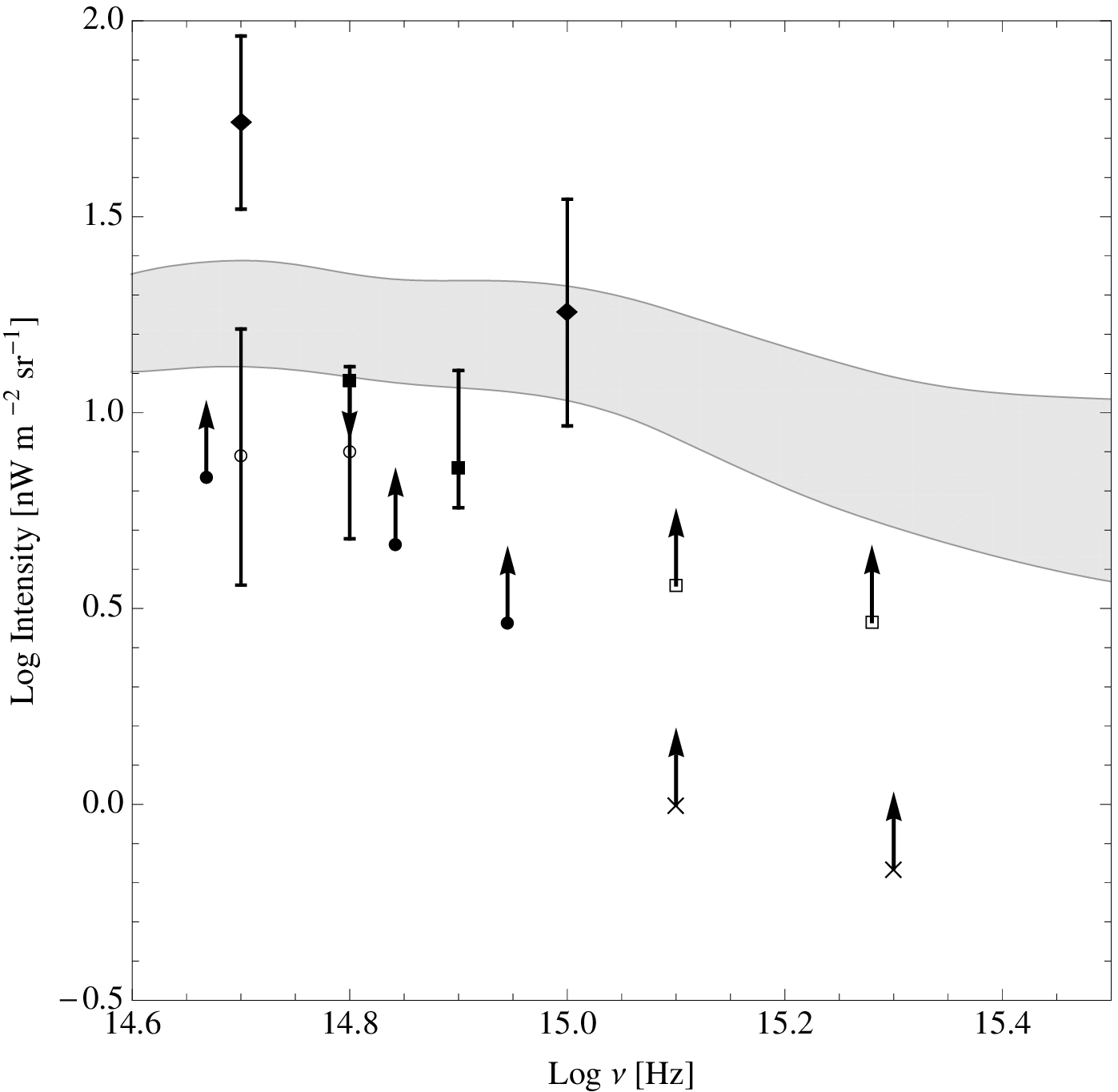}
Figure 3.
\label{EBL}
\end{center}
\end{figure}
\clearpage

\begin{figure}
\begin{center}
\includegraphics[height=7in]{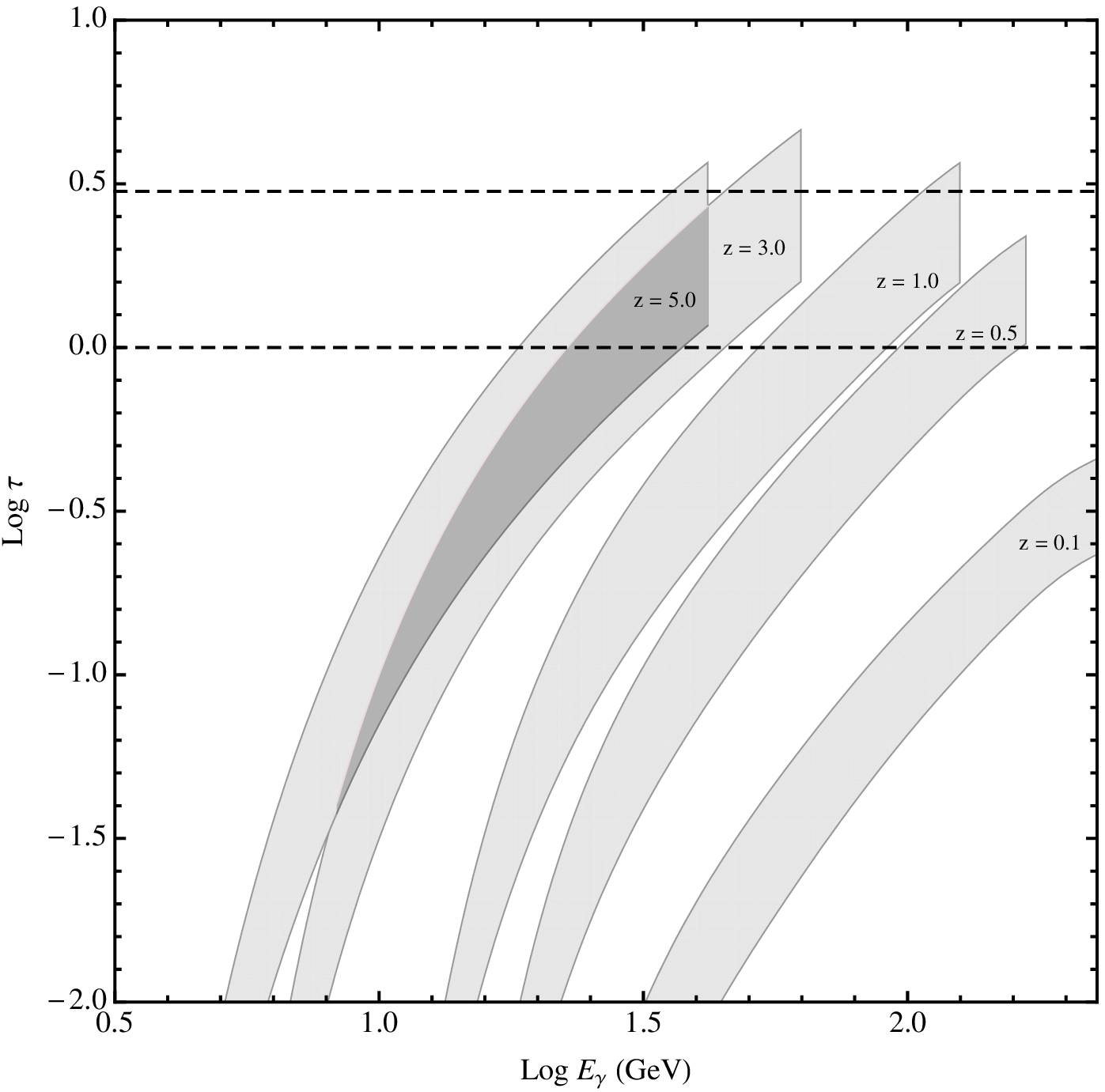}
Figure 4.
\label{opacities}
\end{center}
\end{figure}

\clearpage

\begin{figure}
\begin{center}
\includegraphics[width=7in]{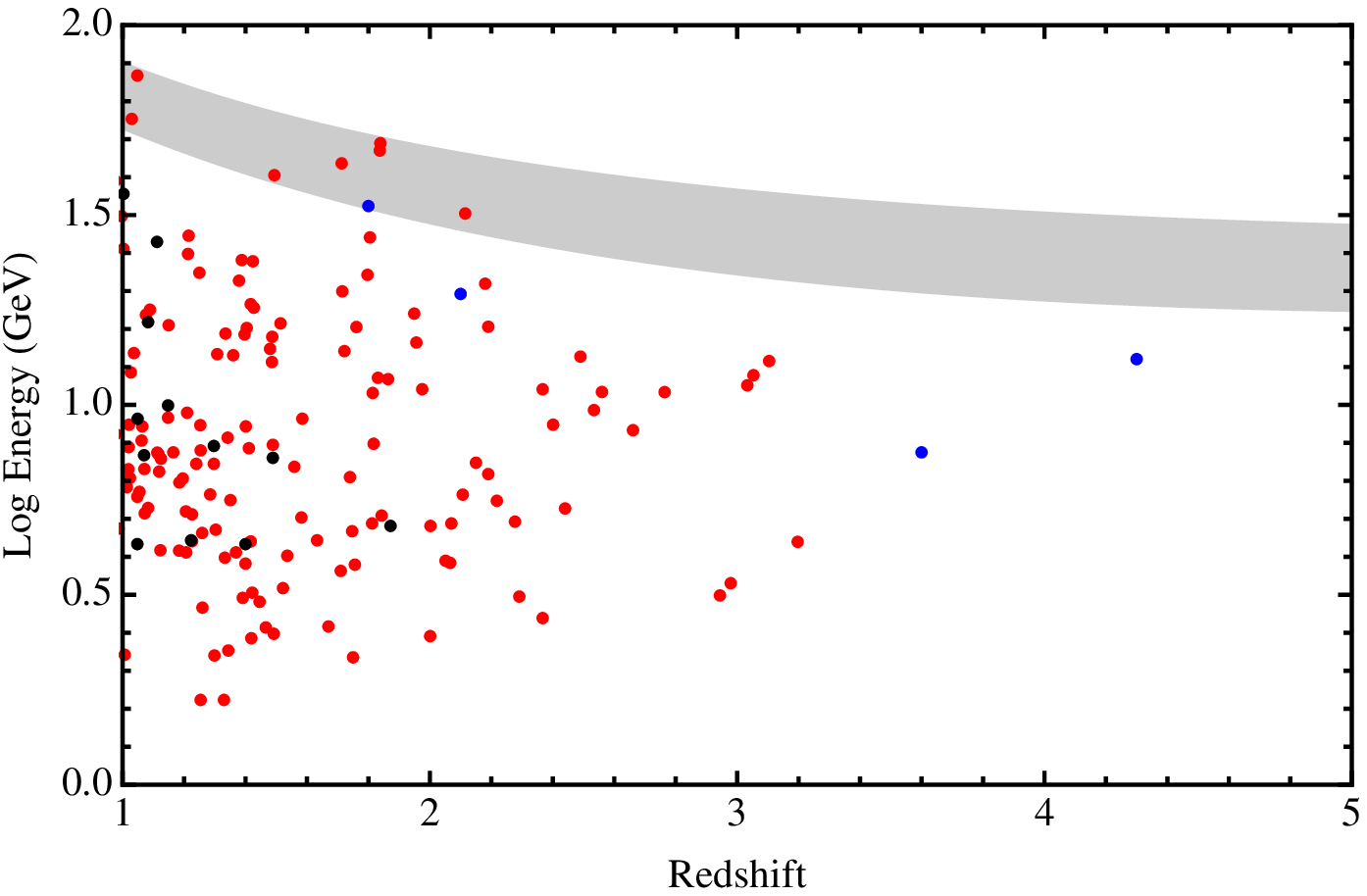}
Figure 5.
\label{FSplot}
\end{center}
\end{figure}

\clearpage

\begin{figure}
\begin{center}
\includegraphics[width=7in]{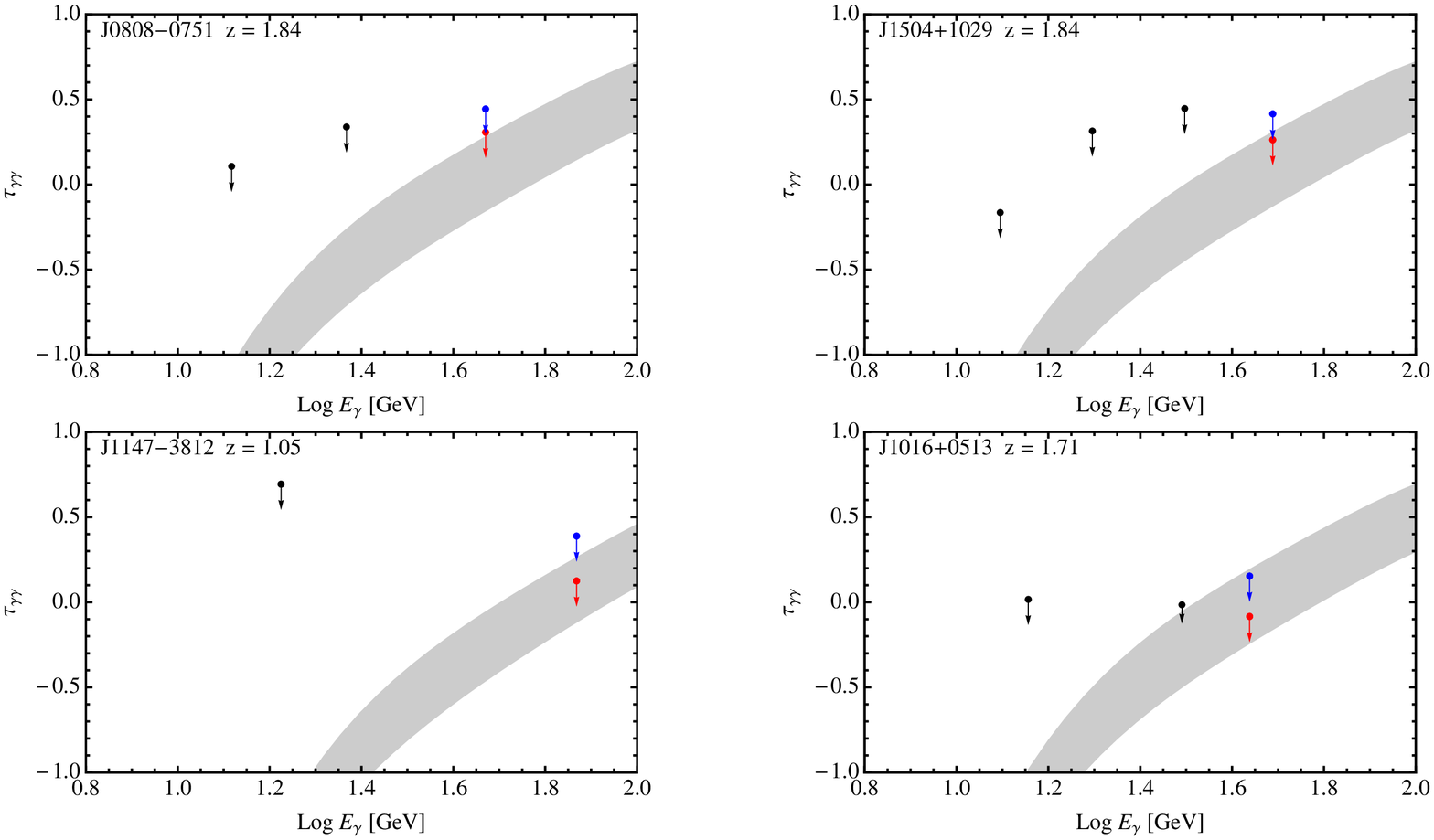}
Figure 6.
\label{Fermilim}
\end{center}
\end{figure}

\end{document}